\documentclass[aps,prb,twocolumn,groupedaddress]{revtex4} 
\usepackage{graphicx}% Include figure files  
\begin{document}  
\newcommand{\be}{\begin{equation}} 
\newcommand{\ee}{\end{equation}} 
\newcommand{\bea}{\begin{eqnarray}} 
\newcommand{\eea}{\end{eqnarray}} 
\newcommand{\nt}{\narrowtext} 
\newcommand{\wt}{\widetext}  

\title{The effect of Fermi surface curvature on low-energy properties of fermions with 
singular interactions} 

\author{A. V. Chubukov$^{1}$ and D. V. Khveshchenko$^{2}$} 

\affiliation{
$^{1}$ Department of Physics, University of Wisconsin, Madison, WI 53706\\
$^{2}$ Department of Physics and Astronomy, University of North Carolina, Chapel Hill, NC 27599}  

\begin{abstract} 
We discuss the effect of Fermi surface curvature on long-distance/time asymptotic behaviors of two-dimensional 
fermions interacting via a gapless mode
described by an effective gauge field-like propagator.
By comparing the predictions based on the idea of multi-dimensional bosonization with 
those of the strong- coupling Eliashberg approach, we demonstrate that 
an agreement between the two requires a further extension of the former technique. 
\end{abstract}  

\maketitle 

In recent years, the behavior of fermions coupled via singular (long-range and/or retarded) 
interactions has been at the forefront of theoretical research in condensed matter physics. 
Such singular interactions are often associated with the ground state instabilities 
and concomitant non-Fermi-liquid behaviors which might occur even in those systems whose microscopic 
Hamiltonians involve only short-range couplings.

In a close 
proximity to the corresponding quantum phase transition, an effective singular coupling is 
mediated by (nearly gapless) collective excitations of the emergent order parameter of either charge or spin nature. 
Important examples include such extensively studied problems as 
antiferromagnetic \cite{abanov,lee} and charge \cite{dicastro} ordering transitions in hole-doped cuprates, 
quantum-critical ferro- \cite{hertz} and antiferro- \cite{chubukov1} magnetic instabilities in heavy 
fermion materials, compressible Quantum Hall Effect (QHE)\cite{lhr}, and Pomeranchuk transitions in low-dimensional
electron gases. The Pomeranchuk transition has been originally discussed in relation to the transport anisotropies reported in QHE
 systems at large half-integer 
filling factors \cite{lilly}. In a more general setting, the idea of a 
spontaneous Pomeranchuk-like distortion of the Fermi surface (FS) 
associated with the transition to a rotationally anisotropic 
"nematic" state in a generic fermion system was put forward in 
Refs.\cite{fradkin1,metzner,kim}.

Despite their different physical nature, the systems studied in 
Refs.\cite{lee,dicastro,hertz,chubukov1,lhr,fradkin1}
conform to the model of a finite density gas of 
two-dimensional fermions coupled via a collective bosonic mode, whose own dynamics 
is described by the (transverse) gauge field-like propagator 
\be
\chi(i\omega,q)=-{\chi_0\over {\gamma|\omega|/q+q^2}}
\label{1}
\ee 
A singular nature of the effective interaction (1) 
manifests itself in singular corrections to the fermion propagator
$G^0(\omega,{\vec k})=1/(i\omega-\xi_k)$, where a generic fermion dispersion 
$\xi_k=v_F {\tilde k}_{\perp}+
\beta {\tilde k}^2_{\parallel}, ~~\tilde {\bf k} = {\bf k} - {\bf k}_F$,
accounts for a non-zero FS curvature of order $\beta\sim v_F/k_F$. 

To first order, the fermion self-energy defined 
as $\Sigma(\omega,{\vec k})=G^{-1}(\omega,{\vec k})-[G^0(\omega,{\vec k})]^{-1}$ takes the form
\cite{lee}
\be
\Sigma_1(i\omega,{\vec k})=\int {d\Omega d{\vec q}\over (2\pi)^3}\chi(i\Omega,{\vec q})G^0
(i\omega+i\Omega,{\vec k}+{\vec q})\sim i\omega_0^{1/3}\omega^{2/3}
\label{3}
\ee
where 
$\omega_0 \sim \chi^3_0/(v^3_F \gamma)$.
At energies $\omega<\omega_0$ the one-loop self-energy (2) exceeds 
the linear in energy term in the bare propagator. Therefore, 
it can no longer be treated as a perturbation, and the higher-order 
contributions must be considered as well.

If one chooses to completely neglect the FS curvature altogether, a 
naive perturbation series expansion for the self-energy appears to produce 
increasingly more and more divergent terms, the $n^{th}$ order term
behaving as $\Sigma_n\propto \omega^{1-n/3}$ 
(Refs. \cite{aim,chubukov2}). 
However, a finite FS curvature provides a regularization of such spurious divergences
\cite{aim,metzner,chubukov2,kopietz3}. Namely, by treating the bosonic mode governed by 
the propagator (1) as a slow subsystem
and invoking the generalized Migdal theorem, one finds that the
vertex corrections are controled by a (inverse) FS curvature and 
appear to be small in powers of $((\ln a)/a)^2$
for large values of the parameter $a = \beta k_F/v_F$.
 
Proceeding along these lines, the authors of ~Refs.\cite{metzner,aim,chubukov2} 
developed a self-consistent, Eliashberg-type, approach, according to which 
the all-order ansatz for the fermion self-energy demonstrates a 
distinctly non-Fermi-liquid behavior, $\Sigma (\omega) \propto \omega^{2/3}$.

As a result, the equal time fermion propagator 
\bea
&&G(0,{\vec r})=\int {d\omega d^2{\vec k}\over (2\pi)^3}{e^{i{\vec k}
{\vec r}}\over {i\omega+i\omega^{1/3}_0\omega^{2/3}-\xi_k}} = 
{\nu_F\over \pi}
\int^\infty_0d\omega \nonumber \\
&& \times \int^\infty_{-\infty}d\xi_k 
{J_0(k_Fr+r\xi_k/v_F)\over {i\omega+i\omega^{1/3}_0\omega^{2/3}-\xi_k}}
\sim G^0 (0,{\vec r}) \left(\frac{r_0}{r}\right)^{1/2}
%(r\chi_0^3)^{1/2}}
\label{5}
\eea
exhibits an algebraic decay which is faster than that of its free fermion counterpart, 
$G^0(0,r) \propto e^{ik_Fr}/(k_Fr^3)^{1/2}$.
The asymptotic behavior (\ref{5}) sets in at distances $r> r_0 \sim \gamma v^4_F/\chi^3_0$.
Albeit being seemingly independent of the FS curvature,
Eq.(\ref{5}) was derived under the condition of convergence of the 
perturbative expansion for the self-energy $\Sigma(\omega)$,
which can only be guaranteed provided that $a = \beta k_F/v_F \gg 1$~\cite{chubukov2}.

In the physically relevant situation ($a \sim 1$) the 
vertex corrections become of order one, and the perturbative expansion ceases to be controllable. 
Nevertheless, it has been conjectured in ~\cite{aim} that at the lowest energies
the results of the Eliashberg theory remain qualitatively applicable,  
and that the power-law behavior of $G(0,r)$ can only be altered in 
the unphysical limit of vanishing FS curvature, $a \rightarrow 0$
 (see also Ref.\cite{metzner2}). 

However, these findings have been recently challenged by the results 
obtained by the method of multi-dimensional bosonization in the 
context of the problem of $2D$ quantum nematic states.
Devised as an alternative to the diagrammatic perturbation theory, 
the early version of multi-dimensional bosonization 
was based on a heuristic idea of dividing the FS onto 
small "patches" and introducing quasi-$1D$ charge (spin) density operators \cite{luther}.
While being sufficient for reproducing the 
correct behavior of the two-particle amplitudes at low energies and 
momenta, such a simplification has not been fully justified in the case of a
single-particle amplitude or even the $2k_F$-behavior of a two-particle one \cite{neto}.

Such an uncertainty notwithstanding, 
the bosonization technique of Refs.\cite{neto} was
applied to the equal time fermion propagator which was found to decay 
faster than any power-law, $G(0,r) \propto\exp({-(r/r_0)^{1/3}})$ \cite{fradkin2}.
The authors of Ref.\cite{fradkin2} argued that that 
this exponential behavior cannot be obtained within the Eliashberg 
theory, thus raising concerns that the latter might be intrinsically incomplete. 
On these grounds, the applicability of the Eliashberg theory was 
also questioned in other contexts, including the high-$T_c$ 
superconductors \cite{frad_kiv}.
   
In view of the continuing controversy, in this Letter we set out to revisit the status of the 
bosonization results pertinent to the gauge-fermion problem. Following the work of Refs.\cite{aim}, 
we focus on the role of the FS curvature which was neglected in Ref.\cite{fradkin2}.

In the well-studied $1D$ case, any deviation from the linear fermion dispersion gives rise 
to cubic terms in the corresponding bosonized theory.
While such terms do spoil the Gaussian form of the bosonic action, they appear to be subdominant,
as far as the asymptotic long-range behavior of the one-and two-fermion amplitudes is concerned. 
When treated self-consistently, though,
these cubic terms produce quadratic corrections to the spectrum of the $1D$ charge and spin density modes \cite{samokhin}.

In $D>1$ dimensions, the situation appears to be more involved, as the (potentially irrelevant) 
quadratic corrections to the fermion spectrum 
$\sim q_n^2=({\vec n}{\vec q})^2$, which depend solely on the component of the transferred 
fermion momentum $\vec q$ parallel to a unit vector $\vec n$ normal to the FS, 
are always complemented by those 
quadratic in terms of the tangential to the FS component, 
$ q_t^2=({\vec n}\times{\vec q})^2$.
The latter terms have no $1D$ analogs and represent a genuine effect of the FS curvature, 
as opposed to a merely non-linear dispersion.
Therefore, they can not be a'priori discarded on the same basis as those associated with 
the normal component of the transferred momentum.

According to the bosonization recipe, a general formula for 
the $2D$ fermion propagator reads \cite{neto}
\be
G(t,{\vec r})=\oint_{FS}{d{\vec n}\over 2\pi}
{d\omega\over 2\pi}{d{\vec q}\over (2\pi)^2}
e^{i({\vec q}{\vec r}-\omega t)}G^0_{\vec n}(\omega,{\vec q})Z_{\vec n}(t,{\vec r})
\label{6}
\ee
where the quasi-$1D$ "patch" Green function 
\be
G^0_{\vec n}(\omega,{\vec q})={1\over {i\omega-v_F{\vec n}{\vec q}}}=\int{dt}{d{\vec r}}
{e^{-i({\vec q}{\vec r}-\omega t)}\over {iv_Ft-{\vec r}{\vec n}}}
\label{7}
\ee
describes $1D$ fermion motion in the direction of the normal vector $\vec n$ defining a FS patch of linear size $\Lambda<<p_F$.

The impact of the interaction on the fermion propagator is encoded in the "eikonal" 
(Debye-Waller-type) factor
$Z_{\vec n}(t,{\vec r})=\exp[-\Phi_{\vec n} (t,{\vec r})]$ where
\bea 
&&\Phi_{\vec n} (t,{\vec r})=\int{d\omega d{\vec q}\over (2\pi)^3}
\chi(\omega,{\vec q})G^0_{\vec n}(\omega,{\vec q})G^0_{\vec n}(-\omega,-{\vec q}) \nonumber \\
&&(1-\cos(\omega t-{\vec r}{\vec q})),
\label{8}
\eea
which expression is common to 
any approximate scheme, where the system of interacting fermions is substituted by 
an effective single-particle environment composed of bosonic collective modes.

With the FS curvature neglected, Eq.(\ref{8}) features a
$1D$ effective interaction $\chi_{1D}(\omega,q_n)=\int({dq_{t}}/2\pi)\chi(\omega,{\vec q})\propto {\omega^{-1/3}}$
between the quasi-$1D$ fermions belonging to a given patch. 
The applicability of the whole scheme hinges on the expectation 
that for a sufficiently singular interaction function, such as that of Eq.(1), 
the scale $\Lambda$ which sets the upper limit in the integration over the transverse 
momentum $q_t$ will effectively drop out of Eq.(\ref{8}). 

However, a simple analysis shows that a typical value of the tangential component of 
the transferred momentum $q_t\sim\omega^{1/3}$ is by far greater than the normal one, $q_n\sim \omega$. 
Therefore, the validity of the assumption about the irrelevance of the FS curvature is anything but granted.

In order to assess the applicability of Eq.(\ref{8})
in the case of a finite FS curvature, we compare it with the direct perturbative expansion for $G(0,r)$. 
To that end, we formally expand 
$Z_{\vec n}(0,r)$  in powers of $\Phi_{\vec n}(0,r)$ and make use of the identity
\be
G^0(\omega,{\vec k})G^0(\omega + \Omega, {\vec k}+{\vec q})={G^0(\omega,{\vec k})-
G^0(\omega + \Omega, {\vec k}+{\vec q})\over {i\Omega - 
\xi_{k+q}+ \xi_k}},
\label{9}
\ee
where $G^0(\omega,{\vec k})$ is the bare fermion Green function.

We explicitly verified that the first order correction given by Eqs.(4) and (6)
can be cast in the equivalent form 
\be
G(0,{\vec r}) = G_0(0,{\vec r}) +  
\int {d\omega d{\vec k}\over (2\pi)^3} 
e^{i{\vec k}{\vec r}}G^2_0(\omega,{\vec k})\Sigma_1(\omega,{\vec k})
\label{10}
\ee
where the lowest-order self-energy is given by Eq.(\ref{3}).
Thus, to first order, 
the bosonization and perturbation theory results agree with each other, and 
the FS curvature does not manifest itself.

However, an explicit comparison between the second order 
term in the expansion of Eq.(\ref{8}) and that of the perturbation theory for the 
self-energy demonstrates that 
the corresponding expressions can only be reconciled provided that one uses
the fermion Green function
\be
G^0_{\vec n}(\omega,{\vec q})={1\over {i\omega-v_F{\vec n}{\vec q}+
\beta({\vec n} \times {\vec q})^2}}
\label{11}
\ee
instead of that with the linearized fermion dispersion. We also found that
functional forms of the higher order cotributions to the self-energy $\Sigma (\omega)$ 
obtained by means of perturbation theory and eikonal approximation 
agree with each other, provided that $G_0$ is given by (\ref{11}), 
although the corresponding prefactors do not necessarily match. 
This discrepancy should have been expected, though.
Indeed, while being able to capture the main effect of small-angle scattering
due to singular interactions, the eikonal approximation is not expected to be exact in $2D$. 

Taken at its face value, the above observation shows a relevance of the FS curvature. It also
suggests a way to improve on the results obtained by virtue of the 
original bosonization technique~\cite{comm_a}. To explore the consequences of 
using Eq.(\ref{11}) for $G^0_{\vec n}(\omega,{\vec q})$, 
we study a spatial dependence of Eq.(\ref{8}) modified in accordance with the above prescription. 

Substituting (\ref{11}) into (\ref{8}) and
introducing $r_n={\vec n}{\vec r}$, we readily obtain 
\be
{\Phi}_{\vec n} (t,{\vec r}) = \frac{\chi_0}{2\pi^3v_F \gamma} 
\int^\infty_0{dq_n\over {\beta q_n}}
(1-\cos q_nr_n)S\left(\frac{v_F q_n}{\gamma^2 \beta^{3}}\right),
\label{12}
\ee
where the integrand reads
\be
S(z)=\int^\infty_0dx\int^\infty_0dy{1\over 
{{\sqrt {zx^3}}+ y}}
Re[{1\over {(1-iy)^2-x^2}}]
\label{13}
\ee
Evaluating the integrals in Eq.(\ref{13}), we find  
that at $z\gg 1$ (e.g., in the limit of a vanishing FS curvature) $S(z)\approx (4\pi^2/27)z^{-1/3}$,
thus yielding
\be
{\Phi}_{\vec n}(0,{\vec r}) \approx \left(\frac{r_n}{r_0}\right)^{1/3}
\label{14}
\ee
where 
$r_0 = [3\sqrt{3}v^{4/3}_F \gamma^{1/3}/(\Gamma(2/3)\chi_0)]^3$.
 This asymptotic behavior agrees with the result obtained in Ref.\cite{fradkin2}.

It is worth mentioning that the function $S(z)$ 
approaches its large-$z$ limit rather slowly, and, therefore, the 
true asymptotic behavior (\ref{14}) sets in at the values of $z$ 
in excess of $\sim 10^{3-4}$ which, in physical applications, lie well outside the realm of any practical interest.

In the opposite limit, $z\ll 1$,
the function (\ref{13}) attains a constant value 
$S(z)\approx(\pi^2/8)$ (Ref. \cite{extra}),
 thereby giving rise to 
the logarithmic behavior
\be
{\Phi}_{\vec n}(0, r_n)\approx\frac{\chi_0}{16\pi\gamma \beta v_F} 
\log({r_n\beta^3\gamma^2\over v_F})   
\label{15}
\ee
Consequently, at the longest distances 
the equal-time fermion propagator shows 
a power-law decay $G(0,{\vec r})\propto 1/r^{\frac{3}{2}+\eta}$ governed by the anomalous exponent 
$\eta=\chi_0/(16\pi\beta \gamma v_F)$. This behavior is in a qualitatively agreement with that predicted by the Eliashberg theory. 

For a quantitative comparison, 
it might be instructive to evaluate the exponent $\eta$ for the parameters of Eq.(1) computed 
(rather than postulated) under the assumption that the dynamics of the collective mode is 
governed by the fermion polarization itself, 
$\chi(\omega,q)\approx\chi_0/(q^2 + \Pi(\omega,q))$~\cite{chubukov2}. In this situation, which tends to be rather common 
in strongly correlated systems, one has $\gamma=m\chi_0/(2\pi v_F)$
with $m = k_F/v_F$. By making an additional assumption of a circular FS, i.e., 
$\beta=1/(2m)$, one obtains $\eta=1/4$, which is only a factor of two short of the Eliashberg result (\ref{5}).

In light of the above, we conclude that
the "minimal" way of accounting for a finite FS curvature through Eq.(\ref{11}) allows one to arrive at a qualitative agreement with the results of Refs.\cite{aim,metzner,chubukov2}, although any further progress towards a quantitative agreement is likely to require an even more drastic rectification of the original bosonization scheme. 

In this regard, our conclusions differ from those drawn in Ref.\cite{kopietz2} where the first attempt to account for the effects of the FS curvature was made. The authors of \cite{kopietz2}
claimed that the FS curvature merely provides a cutoff
for the infrared divergences, so that $Z(0,r_n)$ remains finite  at $r_n\to\infty$, 
and the equal-time fermion propagator retains the free fermion behavior with $\eta=0$. 
We believe that the 
logarithmic divergence (\ref{15}) was overlooked in Ref.\cite{kopietz2}.

For completeness, we also consider the complimentary limit of Eq.(\ref{8}) at large $t$ and $r=0$. 
The corresponding asymptotic behavior of the $\Phi$-factor is then given by the formula
\bea
\Phi_{\vec n}(t,0)={\chi_0\over 2\pi^2v_F}\int^\Lambda_0{d\omega\over {\beta\omega}}
(1-\cos\omega t)\int^\infty_0
{du\over 1+u^3}\nonumber\\
Im[{1\over {\sqrt{u^2+\omega^{2/3}(\omega^{1/3}+i\beta u^2)^2}}}]~~~~
\label{16}
\eea 
At $t<1/\beta^3$,
the curvature is unimportant, and 
 a direct evaluation of Eq.(\ref{16}) gives the following result 
\be
\Phi_{\vec n}(t,0)=-{\Gamma^2({3\over 4})\over 4\pi}{\sqrt {\chi_0\over E_F}}+{1\over 6E_Ft}
\label{17}
\ee
where $E_F=\pi\gamma v^3_F/\chi_0$, which equals $mv^2_F/2$ if the value of $\gamma$
is calculated self-consistently (see above). 

Notably, the expression (\ref{17}) is real, independent of the patch 
size $\Lambda$, and approaches its long-time asymptotic value as $1/t$. 
This is very different from the zero-curvature behavior of 
 $\Phi_{\vec n}(0,{\bf r}) \propto r^{1/3}$.  The difference stems from the fact 
 that finite $r$ and $t$ provide two different regularizations of the 
double Green's function pole in Eq.(\ref{8}). 
Also, contrary to the case of the equal-time behavior, 
the FS curvature affects only the subdominant term in (\ref{17}) 
by replacing it with $\sim (1/t)\ln (t\beta^3\gamma^2)$ for $t>1/(\gamma^2\beta^3)$. 

The difference between $\Phi_{\vec n}(t,0)$ and $\Phi_{\vec n}(0,{\bf r})$
has been previously observed in Refs.\cite{aim} and \cite{fradkin2}. The authors of Ref.\cite{aim} neglected any terms with $E_F$ in the denominator, thus arriving at the result $\Phi_{\vec n}(t,0) =0$.
In contrast, the authors of Ref.\cite{fradkin2} performed the same computation as we did, but found that the 
time-dependent term in $\Phi_{\vec n}(t,0)$ decays 
as $t^{-2/3}\ln t$, which we believe to be the result of a technical error. 

A general failure of the heuristic $D>1$-bosonization technique to properly account for the 
(apparently, important) effects of the FS curvature can be traced back to the substitution 
of the underlying $W_\infty$ algebra of the phase space transformations  
\cite{jevicki} with the approximate $U(1)$/$SU(2)$ Kac-Moody commutation relations 
for the quasi-$1D$ charge/spin density operators \cite{neto}. 
An importance of taking into account the true algebraic structure of the bosonized theory 
was elucidated in the "geometrical bosonization" approach which
strived to reformulate the dynamics of interacting fermions as a purely 
geometric theory of the fluctuating FS \cite{dvk1}.

Notably, a full account of the exact $W_\infty$ algebraic relations has already 
proven to be instrumental for calculating exact correlation 
functions of the $1D$ Calogero-Sutherland model (which also includes the case of non-interacting $1D$ 
fermions with parabolic dispersion) \cite{dvk2}.
This algebraic structure is also present in a more recent reincarnation of 
the idea of geometric bosonization that has been independently developed in the theory 
of mesoscopic transport under the name of "ballistic $\sigma$-model" \cite{efetov}.

In summary, we revisited the problem of two-dimensional fermions coupled to a gauge-field-like collective mode. Comparing the formula for the fermion propagator obtained by means of the multi-dimensional 
bosonization of Refs.\cite{neto} with that found in the framework of the Eliashberg 
approach \cite{aim,chubukov2}, we observed that in order for the two 
to agree, the bosonization prescription must be modified in order to incorporate the FS curvature.
Contrary to the earlier claims, we find that including the FS curvature into Eq.(6) alters 
the predictions of the original bosonization approach, thereby
resulting in a qualitative agreement with the Eliashberg theory.

The authors acknowledge helpful discussions with E. Fradkin, M. Lawler, 
D. Maslov, W. Metzner, and A. Millis.
 This research was supported by NSF-DMR 0240238 (A. V. Ch.) and DMR-0349881 (D. V. K.).


\begin{thebibliography}{99}
\bibitem{extra} A care must be taken when evaluating $S(z)$ at small $z$,
as the limiting value $S(z \rightarrow 0) = \pi^2/8$ differs from $S(z=0) = -\pi^2/4$.
\end{thebibliography}
\end{document}